# Ultrafast light-driven simultaneous excitation of coherent terahertz magnons and phonons in multiferroic BiFeO$_3$

Pritam Khan,[1,2,*] Masataka Kanamaru,[1] Keita Matsumoto,[1,3] Toshimitsu Ito,[4] and Takuya Satoh[1,3,†]

[1] *Department of Physics, Kyushu University, Fukuoka 819-0395, Japan*
[2] *Department of Physics, Bernal Institute, University of Limerick, V94T9PX, Ireland*
[3] *Department of Physics, Tokyo Institute of Technology, Tokyo 152-8551, Japan*
[4] *National Institute of Advanced Industrial Science and Technology, Tsukuba 305-8565, Japan*

The ultrafast switching of magnetization in multiferroic materials by a femtosecond laser could provide various advantages in photonics and magnonics. An efficient approach to control the light-matter interaction is the modulation of ultrafast coherent magnons and phonons in the high-frequency range. Spontaneous Raman and infrared spectra reveal the excitation of magnons and optical phonons in multiferroic BiFeO$_3$ in the sub-few-terahertz range. However, coherent control of such quasiparticles has not been achieved yet. In this study, we demonstrate that linearly polarized laser pulses simultaneously excite coherent magnons [out-of-plane ($\Psi$) and in-plane ($\Phi$) cyclone modes] and optical phonon ($E$ mode) in BiFeO$_3$. Experimental results in conjugation with phenomenological theory, by considering three uniformly distributed magnetic domains, reveal that impulsive stimulated Raman scattering is responsible for the generation of coherent magnons and phonons in BiFeO$_3$. The observation of these terahertz magnon and optical phonon modes paves the way for the development of ultrafast magnetoelectro-optical devices.

[*]pritam.khan@ul.ie
[†]satoh@phys.titech.ac.jp

## I. INTRODUCTION

The multiferroics, unique materials simultaneously exhibiting ferroelectric and (anti)ferromagnetic orders [1–3], are promising candidates for the fabrication of next-generation spintronic devices with added functionalities [4]. Recently, oxide multiferroics have attracted considerable attention owing to the magnetoelectric coupling between the orders by manipulating magnetic properties by electric field [5] and ferroelectric polarization by magnetic field [6]. $BiFeO_3$ stands out as a canonical material in the multiferroic magnetoelectric research because it exhibits ferroelectricity and antiferromagnetism at room temperature [7–11]. Spontaneous Raman [12–17] and infrared (IR) [18–26] spectroscopic measurements on $BiFeO_3$ have revealed the excitation of magnons and phonons. $BiFeO_3$ also exhibits nontrivial light-matter interaction, which yields exotic phenomena, such as visible-light photovoltaic effect [27,28], photoinduced ultrafast magnetic-field-dependent terahertz (THz) emission [29], and giant electrochromic behavior [30]. Recently, ultrafast pump-probe spectroscopy has been successfully used to investigate the excitation of coherent magnons and acoustic phonons in $BiFeO_3$ in the range below 100 GHz [31,32]. Although such observations are quite enticing, excitations of coherent magnon and optical phonon modes in the sub-few-THz regime have not been reported. For practical application, the development of ultrafast magnetoelectro-optical devices requires the generation of optically controlled polarization-sensitive magnon and phonon modes at considerably higher frequencies of the THz regime.

In this study, femtosecond pump-probe spectroscopy was employed to generate high-frequency ultrafast coherent magnons and optical phonon in a $BiFeO_3$ single crystal and to understand their excitation and detection mechanisms. We obtained magnon modes in sub-THz frequency range whereas a phonon mode was observed in few-THz range. The signed amplitude of magnons and phonon exhibit sinusoidal dependence on the pump azimuth angle, which is originated from impulsive stimulated Raman scattering (ISRS). Temperature-dependent pump-probe measurements further indicate that the magnon modes of 530, 560, and 740 GHz (at 300 K) belong to $\Psi_1^{(2)}$, $\Psi_1^{(1)}$, and $\Phi_2^{(1,2)}$ symmetries, respectively.

## II. MATERIAL AND METHODS

### A. Sample fabrication and characteristics

The experiments were performed on a 150-μm-thick single-crystal BiFeO$_3$, grown by using a modified floating-zone method with laser diodes, parallel to the $\hat{\mathbf{z}} \parallel [111]_{pc}$ orientation [33]. At room temperature, BiFeO$_3$ exhibits a rhombohedral *R*3*c* distorted perovskite structure with a magnetic point group of 3*m*. In addition, it exhibits a *G*-type antiferromagnetic order below the Néel temperature of 640 K [34] with a cycloidal spiral structure having a large modulation period of 62 nm [35,36]. BiFeO$_3$ is high-temperature ferroelectric with a Curie temperature of 1100 K and spontaneous electric polarization of approximately 100 μC/cm$^2$ along the *z* direction [37]. We chose a ferroelectric single domain where the electric polarization is oriented along the sample normal (*z* axis). The cycloidal rotation was assumed to be along the $\hat{\mathbf{x}} \parallel [10\bar{1}]_{pc}$ direction, whereas $\hat{\mathbf{y}} \parallel [\bar{1}2\bar{1}]_{pc}$ was perpendicular to the plane of cycloidal rotation.

### B. Ultrafast pump-probe measurements

To observe the ultrafast spin and lattice dynamics in BiFeO$_3$, we performed femtosecond time-resolved pump-probe measurements in the transmission geometry. A pump pulse generated by a Ti:sapphire laser with a regenerative amplifier and optical parametric amplifier (Spectra-Physics) at a repetition rate of 500 Hz, wavelength of 1300 nm, and duration of 70 fs was incident at an angle of 10° with respect to the sample normal ($z$ axis). The pump photon energy is below the lowest $Fe^{3+}$ $d$–$d$ or $p$–$d$ transition energy [38–40]. Probe pulses with a wavelength of 800 nm and repetition rate of 1 kHz were incident perpendicularly to the sample. The linearly polarized pump beam was focused to a spot diameter of approximately 100 µm, which yielded a fluence of approximately 25 mJ/cm². The circularly polarized probe beam had approximately twice smaller spot diameter. A schematic of the pump-probe measurement and sample orientation are shown in Fig. 1. In the pump-probe measurement, optically excited coherent magnons and phonons were observed by using the polarimetric detection technique [41–43] with balanced detectors enabling the extraction of the contributions of an induced change in ellipticity of the probe polarization. The pump azimuth angle $\theta$ was defined as the angle between the $x$ axis and the direction of the pump polarization (electric field). The detection was performed with a circularly polarized probe in two different geometric configurations, *X*- and *Y*-probe configurations [42], where the detection angles were set to $\rho = 0$ and $\pi/4$, respectively.

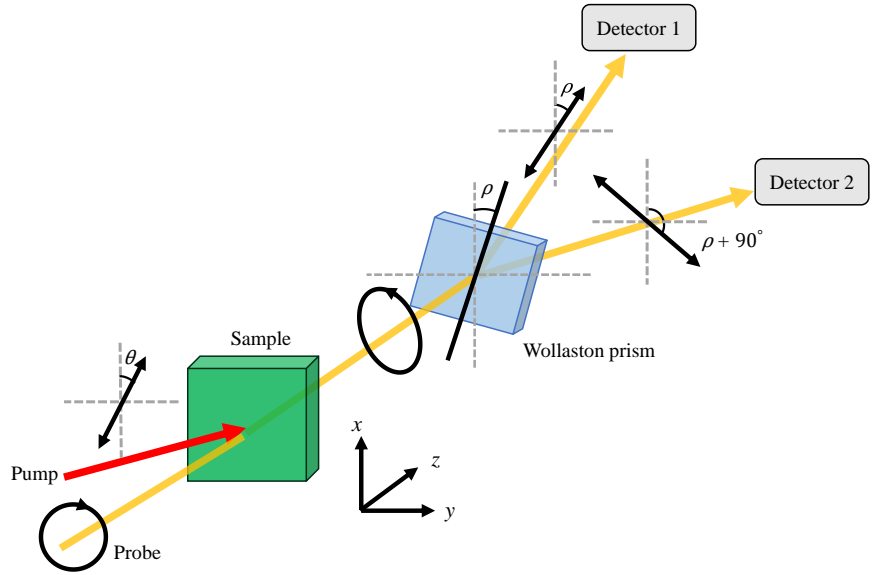

FIG. 1. Schematic of sample orientation pump-probe measurement with a linearly polarized pump and circularly polarized probe. Here $\theta$ and $\rho$ are the pump azimuth and the detection angle, respectively.

## III. RESULTS and DISCUSSION

Figure 2 shows the time evolution of the ellipticity change ($\Delta\eta$) in probe polarization in BiFeO$_3$ at 300 K, for an excitation at a pump azimuth angle of $\theta = -\pi/4$ and detection in the *Y*-probe configuration. We observed oscillatory signals having different characteristic frequencies, as shown in Figs. 2(a) and 2(b), with an abruptness at delay $t \sim 2.8$ ps. The abrupt signal is observed because of the overlap of the probe pulse with the pump pulse, which is reflected from the back surface of the sample, counterpropagated, and reflected again from the front surface [41]. We performed fast Fourier transform (FFT) of the oscillatory signals at positive delays. The resulting amplitude spectra are shown in the insets of Figs. 2(a) and 2(b). The FFT spectra reveal a high-amplitude mode at 2.4 THz in Fig. 2(a) and three low-amplitude modes at 530, 560, and 740 GHz in Fig. 2(b).

Experimental results in conjugation with the symmetry consideration reveal that the ISRS is responsible for the excitation of the 2.4-THz phonon mode having *E* symmetry, which is consistent with the mode in previous reports observed by spontaneous Raman [16] and IR [18] spectroscopies. Detailed results and analysis on the selection rule for the excitation and detection of the coherent phonon are presented in Figs. S1 and S2 of Supplementary Material (SM) [44] (see also Refs. [16,45] therein).

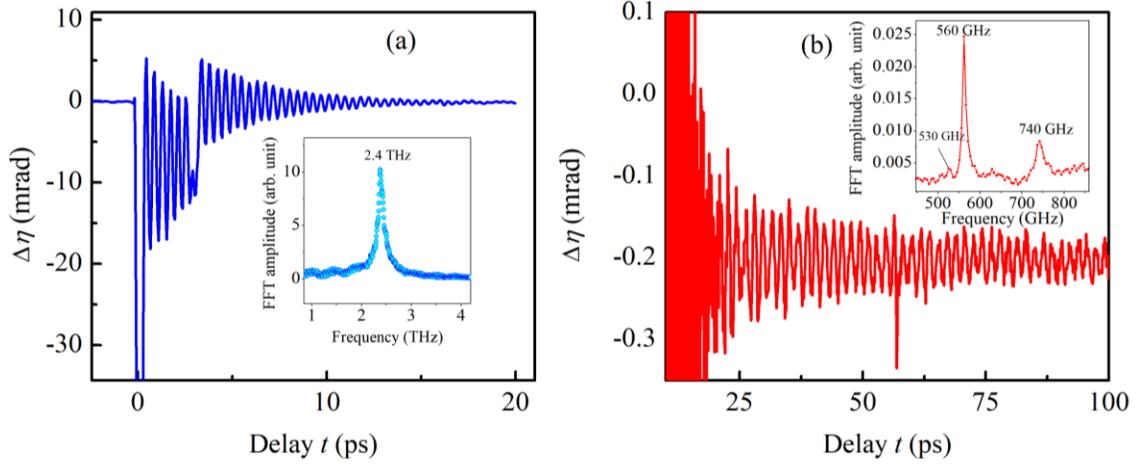

FIG. 2. Temporal evolutions of the ellipticity changes (Δη) in transmitted probe polarizations in BiFeO$_3$ ($T$ = 300 K) at two different probe delays up to (a) 20 and (b) 60 ps. The insets show the FFT amplitude spectra of the signals at the corresponding delays, which reveal center frequencies of (a) 2.4 THz and (b) 530, 560, and 740 GHz.

The temperature dependence of the 2.4-THz frequency is shown in Fig. 3. The FFT spectra of the oscillation curves corresponding to the 2.4-THz mode are shown in Fig. 3(a). The ellipticity curves associated with THz mode is shown in Fig. S3 of SM [44]. Figure 3(b) reveals that the 2.4-THz phonon mode exhibits frequency softening (~ 4%) within the full width at half maximum (FWHM) of the FFT spectra shown in Fig. 3(a), when the temperature is increased.

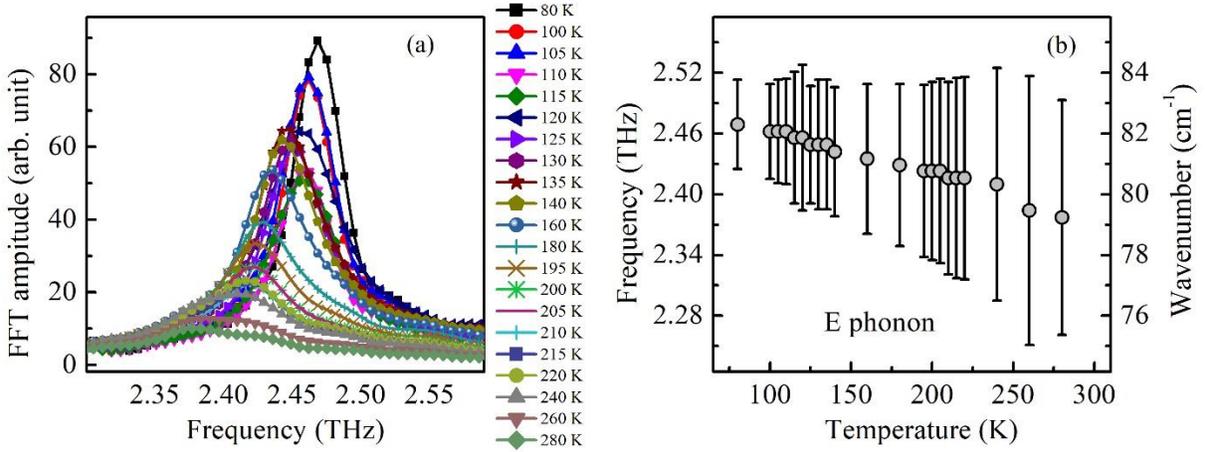

FIG. 3. (a) FFT amplitude spectra of the 2.4-THz mode at various temperatures. (b) Variation of *E* phonon (2.4 THz) at different temperatures. Here, the black line represents the FWHM of the phonon mode.

To understand the temperature dependence of sub-THz modes, ellipticity changes are recorded at various temperatures as shown in Fig. S4 of SM [44]. The corresponding FFT spectra of the sub-THz frequencies at different temperatures are shown in Fig. 4(a). The amplitude of the 530-GHz mode is small and overshadowed by the 560- and 750-GHz mode to be observed. In a stark contrast to 2.4-THz phonon mode, sub-THz modes exhibit stronger temperature dependences (the frequencies decrease drastically with the increase in temperature), as shown in Fig. 4(b) within the FWHM of the FFT spectra in Fig. 4(a). The comparison of the temperature dependences of the frequencies to those of the previously reported IR transmission spectra of BiFeO$_3$ [22–26] shows that the 530-, 560-, and 740-GHz modes are consistent with the $\Psi_1^{(2)}$, $\Psi_1^{(1)}$, and $\Phi_2^{(1,2)}$ magnon modes, respectively.

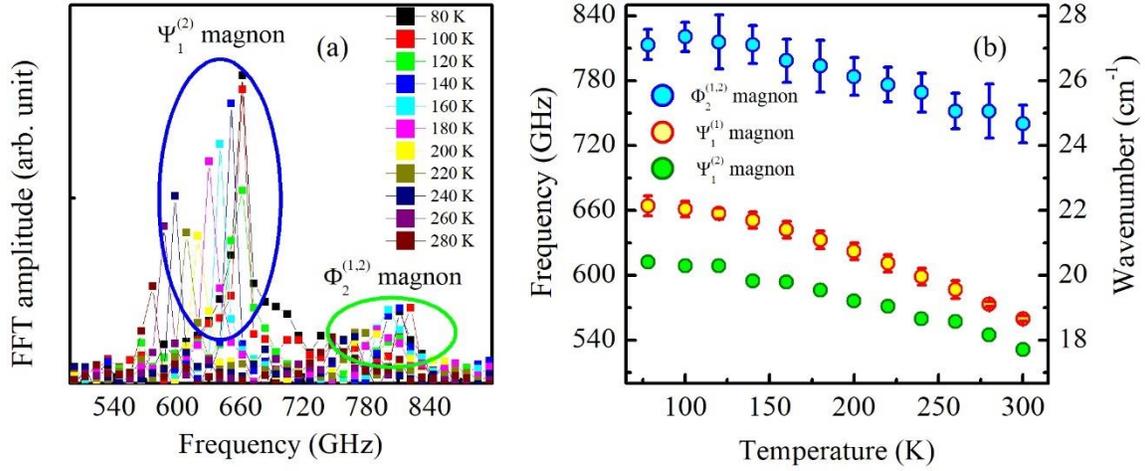

FIG. 4. (a) FFT amplitude spectra of the sub-THz oscillations at various temperatures. (b) Variations of $\Psi_1^{(2)}$ (530 GHz), $\Psi_1^{(1)}$ (560 GHz), and $\Phi_2^{(1,2)}$ (740 GHz) magnon modes as functions of the measurement temperature. Here, the blue and red lines represent the FWHM of the $\Phi_2^{(1,2)}$ and $\Psi_1^{(1)}$ magnon modes, respectively.

To elucidate the selection rule for the excitation and detection of coherent magnons, we performed time-resolved measurements in *X*- and *Y*-probe configurations at various pump azimuth angles $\theta$, as shown in Fig. S5 of SM [44]. Figures 5(a) and 5(b) show the signed amplitudes of the $\Psi_1^{(1)}$ and $\Phi_2^{(1,2)}$ modes in the *X*- and *Y*-probe configurations, respectively. Both modes exhibit the same dependence on the pump azimuth [i.e., $\cos(2\theta)$ and $\sin(2\theta)$ in the *X*- and *Y*-probe configurations, respectively]. As stated above, the amplitude of the $\Psi_1^{(2)}$ mode is small. However, a FFT analysis at selected pump azimuth reveals the appearance of this mode at $\theta = 0$ and $\pi/2$ in the *X*-probe configuration and at $\theta = \pm\pi/4$ in the *Y*-probe configuration as shown in Figs. 5(c) and 5(d). Therefore, the selection rule for the $\Psi_1^{(2)}$ mode is similar to those for the $\Psi_1^{(1)}$ and $\Phi_2^{(1,2)}$ modes.

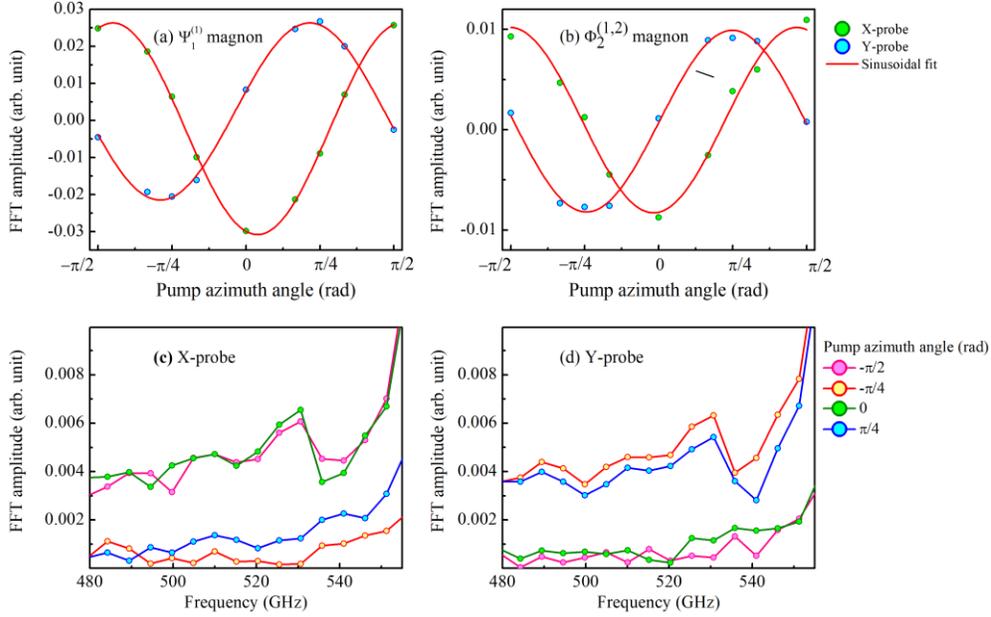

FIG. 5. Dependences of the signed amplitudes of the (a) $\Psi_1^{(1)}$ and (b) $\Phi_2^{(1,2)}$ magnon modes measured in the *X*-probe (green circles) and *Y*-probe (blue circles) configurations as functions of the pump azimuth angle (*T* = 300 K). The solid red curves represent the sinusoidal fits of the experimental data. FFT amplitude spectra of 530-GHz $\Psi_1^{(2)}$ magnon mode at selected pump azimuth angles in the (c) *X*- and (d) *Y*-probe configurations.

Further, we use the phenomenological theory of ISRS to correlate the experimental results for BiFeO$_3$. In principle, the light-matter interaction can be expressed by the interaction free-energy density [46,47],

$$F = -\frac{1}{16\pi}\epsilon_{ij}(\mathbf{M},\mathbf{L})\mathcal{E}_i\mathcal{E}_j^*, \qquad (1)$$

where $\epsilon_{ij}(\mathbf{M},\mathbf{L})$ is the dielectric tensor, a complex function of $\mathbf{M}$ and $\mathbf{L}$ (we define a ferromagnetic vector $\mathbf{M} = \mathbf{M}_1 + \mathbf{M}_2$ and antiferromagnetic vector $\mathbf{L} = \mathbf{M}_1 - \mathbf{M}_2$ by using the two sublattice magnetizations $\mathbf{M}_1$ and $\mathbf{M}_2$), and $\mathcal{E}_i$ is the time-dependent amplitude of the electric field $E_i(t) = \mathrm{Re}[\mathcal{E}_i(t)\exp(i\omega t)]$.

First, we expand the dielectric tensor $\epsilon_{ij}(\mathbf{M},\mathbf{L})$ into a power series of $\mathbf{M}$ and $\mathbf{L}$ [46,48],

$$\epsilon_{ij} = \epsilon_{ij}^{(0)} + ik_{ijk}^M M_k + ik_{ijk}^L L_k + g_{ijkl}^{MM} M_k M_l + g_{ijkl}^{LL} L_k L_l + p_{ijkl}^{ML} M_k L_l, \qquad (2)$$

where $k_{ijk}$ and $g_{ijkl}$ are the magneto-optical coupling constants, which mediate the inverse Faraday effect and inverse Cotton-Mouton effect (ICME), respectively [46]. As $|\mathbf{M}| \ll |\mathbf{L}|$, we can neglect the quadratic term of $\mathbf{M}$, i.e., the fourth term in the right-hand side of Eq. (2). Considering the $3m$ symmetry, the components of $k^M$, $k^L$, $g^{LL}$, and $p^{ML}$ are shown in SM [44] (see also Ref. [49] therein). The antiferromagnetic vector $\mathbf{L}$ exhibits a cycloid rotation in the $xz$ plane [35]. For simplicity, we separately consider the $x$ and $z$ components of $\mathbf{L}$. We also neglect local ferromagnetic components [50]. We start with $\mathbf{L} = (L_x, 0, 0)$, and thus the components of $\epsilon_{ij}(\mathbf{M}, \mathbf{L})$ can be expressed as in Table I.

TABLE I. Dielectric tensor components of BiFeO$_3$ for $\mathbf{L} = (L_x, 0, 0)$.

| Tensor element | $\Psi$ mode ($m_z, l_y \neq 0$) | $\Phi$ mode ($m_y, l_z \neq 0$) |
| --- | --- | --- |
| $\epsilon_{xx}$ | $p_3 L_x m_z$ | $p_4 L_x m_y$ |
| $\epsilon_{yy}$ | $-p_3 L_x m_z$ | $-p_2 L_x m_y$ |
| $\epsilon_{xy}$ | $g_9 L_x l_y + i k_2 m_z$ | $g_4 L_x l_z$ |
| $\epsilon_{yx}$ | $g_9 L_x l_y - i k_2 m_z$ | $g_4 L_x l_z$ |

In the table, $\mathbf{m}$ and $\mathbf{l}$ are the dynamic components of $\mathbf{M}$ and $\mathbf{L}$, respectively. The out-of-plane ($\Psi$) and in-plane ($\Phi$) cyclone modes have $m_z, l_y$ and $m_y, l_z$ nonzero dynamic components, respectively [36,51,52].

We define the effective magnetic fields for $\mathbf{m}$ and $\mathbf{l}$ as

$$\mathbf{H}^{\text{eff}} = -\frac{\partial F}{\partial \mathbf{m}}, \tag{3}$$

$$\mathbf{h}^{\text{eff}} = -\frac{\partial F}{\partial \mathbf{l}} \tag{4}$$

respectively. We define $\mathcal{E}_x(t) = \mathcal{E}_0(t)\cos\theta$ and $\mathcal{E}_y(t) = \mathcal{E}_0(t)\sin\theta \exp(i\varphi)$ ( $0 \leq \theta < \pi$, $-\frac{\pi}{2} \leq \varphi < \frac{\pi}{2}$). As the pulse widths of the femtosecond lasers are significantly smaller than the period of the magnon oscillations, $[\mathcal{E}_0(t)]^2 \to I_0 \delta(t)$, where $I_0 \equiv \int [\mathcal{E}_0(t)]^2 dt$. By using Table I, we can express the components of $\mathbf{H}^{\text{eff}}$ and $\mathbf{h}^{\text{eff}}$ as

$$H_x^{\text{eff}} = 0, \tag{5}$$

$$H_y^{\text{eff}} = \frac{1}{16\pi} I_0 \delta(t)(p_4 L_x \cos^2\theta - p_2 L_x \sin^2\theta), \tag{6}$$

$$H_z^{\text{eff}} = \frac{1}{16\pi} I_0 \delta(t)\{p_3 L_x \cos(2\theta) + k_2 \sin\varphi \sin(2\theta)\}, \tag{7}$$

$$h_x^{\text{eff}} = 0, \tag{8}$$

$$h_y^{\text{eff}} = \frac{1}{16\pi} I_0 \delta(t) g_9 L_x \cos\varphi \sin(2\theta), \tag{9}$$

$$h_z^{\text{eff}} = \frac{1}{16\pi} I_0 \delta(t) g_4 L_x \cos\varphi \sin(2\theta). \tag{10}$$

The equations of motions of $\mathbf{m}$ and $\mathbf{l}$ are [53]

$$\frac{d\mathbf{m}(t)}{dt} = -\frac{1}{2}\gamma(\mathbf{M} \times \mathbf{H}^{\text{eff}} + \mathbf{L} \times \mathbf{h}^{\text{eff}}), \tag{11}$$

$$\frac{d\mathbf{l}(t)}{dt} = -\frac{1}{2}\gamma(\mathbf{M} \times \mathbf{h}^{\text{eff}} + \mathbf{L} \times \mathbf{H}^{\text{eff}}). \tag{12}$$

We neglect the damping terms because they are not important in the analysis. Further, we integrate Eqs. (11) and (12) around $t = 0$, which yields

$$\Delta\mathbf{m} = \frac{1}{32\pi}\gamma I_0 L_x^2 [g_4 \cos\varphi \sin(2\theta)\hat{y} - g_9 \cos\varphi \sin(2\theta)\hat{z}], \tag{13}$$

$$\Delta\mathbf{l} = \frac{1}{32\pi}\gamma I_0 \left[\{p_3 L_x^2 \cos(2\theta) + k_2 L_x \sin(2\theta)\sin\varphi\}\hat{y} + \left\{p_9 L_x^2 \cos(2\theta) - \frac{p_2-p_4}{2}L_x^2\right\}\hat{z}\right]. \tag{14}$$

For a linearly polarized light, $\varphi = 0$, and thus Eqs. (13) and (14) can be simplified to

$$\Delta\mathbf{m}^{\text{lin}} = \frac{1}{32\pi}\gamma I_0 L_x^2 [g_4 \sin(2\theta)\hat{y} - g_9 \sin(2\theta)\hat{z}], \tag{15}$$

$$\Delta\mathbf{l}^{\text{lin}} = \frac{1}{32\pi}\gamma I_0 \left[p_3 L_x^2 \cos(2\theta)\hat{y} + \left\{p_9 L_x^2 \cos(2\theta) - \frac{p_2-p_4}{2}L_x^2\right\}\hat{z}\right]. \tag{16}$$

We define exchange enhancement factors $A_\Psi = \frac{l_y}{m_z}$ and $A_\Phi = \frac{l_z}{m_y}$; usually, $A_\Psi$ and $A_\Phi \gg 1$ [46,47,54]. At $t = 0$, $m_y$ and $m_z$ have finite values, and thus are proportional to $\cos(\omega_\Psi t)$ and $\cos(\omega_\Phi t)$, respectively, whereas $l_y$ and $l_z$ oscillate with a phase difference of $\pi/2$ and exhibit $\sin(\omega_\Psi t)$ and $\sin(\omega_\Phi t)$ dependences, respectively,

$$m_y^{\text{lin}}(t) = \frac{1}{32\pi}\gamma I_0 g_4 L_x^2 \sin(2\theta)\cos(\omega_\Phi t), \tag{17}$$

$$m_z^{\text{lin}}(t) = -\frac{1}{32\pi}\gamma I_0 g_9 L_x^2 \sin(2\theta)\cos(\omega_\Psi t), \tag{18}$$

$$l_y^{\text{lin}}(t) = \frac{1}{32\pi}\gamma I_0 A_\Psi g_9 L_x^2 \sin(2\theta)\sin(\omega_\Psi t), \tag{19}$$

$$l_z^{\text{lin}}(t) = \frac{1}{32\pi}\gamma I_0 A_\Phi g_4 L_x^2 \sin(2\theta)\sin(\omega_\Phi t). \tag{20}$$

As the coupling between the light and magnetization is described solely by the tensor $g$, we confirm that both modes are originated from the ICME. The changes in probe polarization, i.e., ellipticity ($\Delta\eta$), can be expressed as

$$\Delta\eta_X \propto \epsilon_{xx} - \epsilon_{yy} \propto p_3 L_x m_z + p_9 L_x m_y$$
$$\propto p_3 g_9 L_x^3 \sin(2\theta)\cos(\omega_\Psi t) + p_9 g_4 L_x^3 \sin(2\theta)\cos(\omega_\Phi t) \sim (A_{\Psi,\Phi})^0, \tag{21}$$

$$\Delta\eta_Y \propto \epsilon_{xy} + \epsilon_{yx} \propto g_9 L_x l_y + g_4 L_x l_z$$
$$\propto A_\Psi(g_9)^2 L_x^3 \sin(2\theta)\sin(\omega_\Psi t) + A_\Phi(g_4)^2 L_x^3 \sin(2\theta)\sin(\omega_\Phi t). \tag{22}$$

Further, we consider the second case for $\mathbf{L} = (0, 0, L_z)$. Similarly (the calculations are shown in SM [44]), under the conditions $A_\Psi$ and $A_\Phi \gg 1$, the components of $\mathbf{m}$ and $\mathbf{l}$ can be expressed as

$$m_x^{\text{lin}}(t) = \frac{1}{32\pi}\gamma I_0 g_4 L_z^2 \cos(2\theta)\cos(\omega_\Psi t), \tag{23}$$

$$m_y^{\text{lin}}(t) = -\frac{1}{32\pi}\gamma I_0 g_4 L_z^2 \sin(2\theta)\cos(\omega_\Phi t), \tag{24}$$

$$l_x^{\text{lin}}(t) = \frac{1}{32\pi}\gamma I_0 A_\Phi g_4 L_z^2 \sin(2\theta)\sin(\omega_\Phi t), \tag{25}$$

$$l_y^{\text{lin}}(t) = \frac{1}{32\pi}\gamma I_0 A_\Psi g_4 L_z^2 \cos(2\theta)\sin(\omega_\Psi t). \tag{26}$$

Likewise, we obtain the changes in probe polarization,

$$\Delta\eta_X \propto g_4 L_z l_y + p_1 L_z m_x$$

$$\propto A_\Psi(g_4)^2 L_z^3 \cos(2\theta)\sin(\omega_\Psi t) + p_1 g_4 L_z^3 \cos(2\theta)\cos(\omega_\Psi t)$$

$$\approx A_\Psi(g_4)^2 L_z^3 \cos(2\theta)\sin(\omega_\Psi t), \tag{27}$$

$$\Delta\eta_Y \propto g_4 L_z l_x - p_1 L_z m_y$$

$$\propto A_\Phi(g_4)^2 L_z^3 \sin(2\theta)\sin(\omega_\Phi t) - p_1 g_4 L_z^3 \sin(2\theta)\cos(\omega_\Phi t)$$

$$\approx A_\Phi(g_4)^2 L_z^3 \sin(2\theta)\sin(\omega_\Phi t). \tag{28}$$

By combining Eqs. (21), (22) and (27), (28) and considering the contributions of $L_x$ and $L_z$, we obtain the selection rules for the magnon modes, as shown in Table II.

TABLE II. Selection rules for the excitations of coherent magnons in the *X*- and *Y*-probe configurations.

| Probe | Ψ | Φ |
|---|---|---|
| X | cos(2θ) |  |
| Y | sin(2θ) | sin(2θ) |

According to Table II, for the Ψ modes, the experimental results well match with the theoretical calculations in both *X*- and *Y*-probe configurations. However, the experimental and theoretical results for the Φ mode in the *X*-probe configuration contradict.

In the above analysis, we assumed that BiFeO$_3$ has a single magnetic domain with a spin cycloidal rotation along the $\hat{x} \parallel [10\bar{1}]_{\text{pc}}$ direction. Next, we assume that BiFeO$_3$ has three uniformly distributed domains where the spin cycloidal rotations are along the $[10\bar{1}]_{\text{pc}}$, $[\bar{1}10]_{\text{pc}}$, and $[0\bar{1}1]_{\text{pc}}$ directions [55]. A detailed description with the contribution of all three domains and probe light polarization changes by $L_x$ and $L_z$ under the condition $A_\Psi$ and $A_\Phi \gg 1$ are presented in SM [44], while the resulting selection rules are shown in Table III.

TABLE III. Selection rules for the excitations of coherent magnons in the *X*- and *Y*-probe configurations assuming the contributions of multidomains.

| Probe | $\Psi$ | $\Phi$ |
|---|---|---|
| X | $\cos(2\theta)$ | $\cos(2\theta)$ |
| Y | $\sin(2\theta)$ | $\sin(2\theta)$ |

Thus, the consideration of the contributions of all three domains leads to experimental results well matched with the theoretical calculations.

## IV. CONCLUSIONS

The dynamics of coherent magnon and phonon modes in the multiferroic $BiFeO_3$ were investigated by time-resolved ellipticity measurements. Spin and lattice oscillations in $BiFeO_3$ were excited nonthermally by the linearly polarized light. The subsequent detection by the circularly polarized probe, along with the symmetry analysis by considering the contribution of three uniformly distributed magnetic domains, indicated that the ISRS was responsible for the driving of the out-of-plane ($\Psi_1^{(2)}$, $\Psi_1^{(1)}$) and in-plane ($\Phi_2^{(1,2)}$) cyclone magnon modes with frequencies of 530, 560, and 740 GHz at 300 K, respectively. These results pave the way for further studies on electromagnon excitations under the influence of electric and magnetic fields, which could reveal the inherent relationship between the ferroelectric and antiferromagnetic orders in THz optomagnonics.

## ACKNOWLEDGMENTS

We thank S. Miyahara, A. Koreeda, Y. Fujii, T. Yoshimura, and N. Fujimura for valuable discussions and M. Fiebig and Ch. Tzschaschel for helping in the low-temperature pump-probe measurements. This study was supported by the Japan Society for the Promotion of Science (JSPS) KAKENHI (Grants No. JP15H05454, No. JP16F16358, No. JP19H01828, No. JP19H05618, No. JP19J21797, No. JP19K21854, and No. JP26103004), JSPS Core-to-Core Program (A. Advanced Research Networks).

*Supplementary Material*

## Mechanisms of excitation and detection of the 2.4-THz E phonon mode

To determine the selection rule for the excitation and detection of the phonon mode, we consider that BiFeO$_3$ belongs to the 3m point group. Therefore, the possible symmetry modes are A$_1$ and E expressed in Supplementary Table S1 [43].

**Table S1**. Raman tensor of the phonon modes in BiFeO$_3$ associated with the A$_1$ and E symmetry modes. $\hat{x} \parallel [10\bar{1}]_{pc}$, $\hat{y} \parallel [\bar{1}2\bar{1}]_{pc}$, $\hat{z} \parallel [111]_{pc}$. The mirror plane is perpendicular to the $x$ axis.

| A$_1$ | E($y$) | E($-x$) |
|---|---|---|
| $\begin{pmatrix} a & 0 & 0 \\ 0 & a & 0 \\ 0 & 0 & b \end{pmatrix}$ | $\begin{pmatrix} c & 0 & 0 \\ 0 & -c & d \\ 0 & d & 0 \end{pmatrix}$ | $\begin{pmatrix} 0 & -c & -d \\ -c & 0 & 0 \\ -d & 0 & 0 \end{pmatrix}$ |

According to the spontaneous Raman spectra [15], the 2.4-THz mode belongs to the E symmetry. Thus, we focus only on the Raman tensor with the E symmetry to identify the excitation mechanism of this coherent phonon mode by femtosecond laser pulses by choosing the distinct polarization of the pump pulses. In this regard, we need to calculate the driving force of the coherent phonon modes, which depends on the electric field of the pump pulses, $E_i(t) = \mathrm{Re}[\mathcal{E}_i(t) \exp(i\omega t)]$. Assuming light propagates in the $z$ direction ($\mathcal{E} = (\mathcal{E}_x, \mathcal{E}_y, 0)$), the classical equation of motion of a lattice field can be expressed as:

$$\frac{d^2 Q}{dt^2} + \Omega^2 Q = F(t), \tag{S1}$$

where $Q$ is the classical phonon field with an angular frequency of $\Omega$. For ISRS, the driving force is $F = \frac{1}{2} R_{ij} \mathcal{E}_i \mathcal{E}_j^*$, where $R_{ij}$ is the Raman tensor of the sample. For a linearly polarized pump pulse, we define $\mathcal{E}_x(t) = \mathcal{E}_0(t) \cos\theta$ and $\mathcal{E}_y(t) = \mathcal{E}_0(t) \sin\theta$. As the pulse width of the femtosecond laser is significantly smaller than the period of the phonon oscillations, $[\mathcal{E}_0(t)]^2 \to I_0\delta(t)$, where $I_0 \equiv \int [\mathcal{E}_0(t)]^2 dt$.

Consequently, the driving force for the E phonon mode can be expressed as

$$F[\text{E}(y)] = c[\mathcal{E}_x^2 - \mathcal{E}_y^2] = cI_0\delta(t)\cos(2\theta), \tag{S2}$$

$$F[\text{E}(-x)] = -2c[\mathcal{E}_x\mathcal{E}_y] = -2cI_0\delta(t)\sin(2\theta). \tag{S3}$$

The Raman tensor matrix and driving force calculation show that the $\text{E}(y)$ mode has no effective off-diagonal components. Hence, the detection of coherent phonons requires a detection angle of $\rho = -\pi/4$ or $\pi/4$. In contrast, the $\text{E}(-x)$ mode has no diagonal components. Hence, we set $\rho = -\pi/2$ or $0$ for the detection of coherent phonons. Therefore, to excite and detect the coherent E phonons with a frequency of 2.4 THz, we performed pump–probe measurements in two different experimental settings (Fig. S1), X- and Y-probe configurations, associated with the selective excitation and detection of the $\text{E}(y)$ and $\text{E}(-x)$ modes. Fig. S1 shows that experimental results match well with the theoretical calculation.

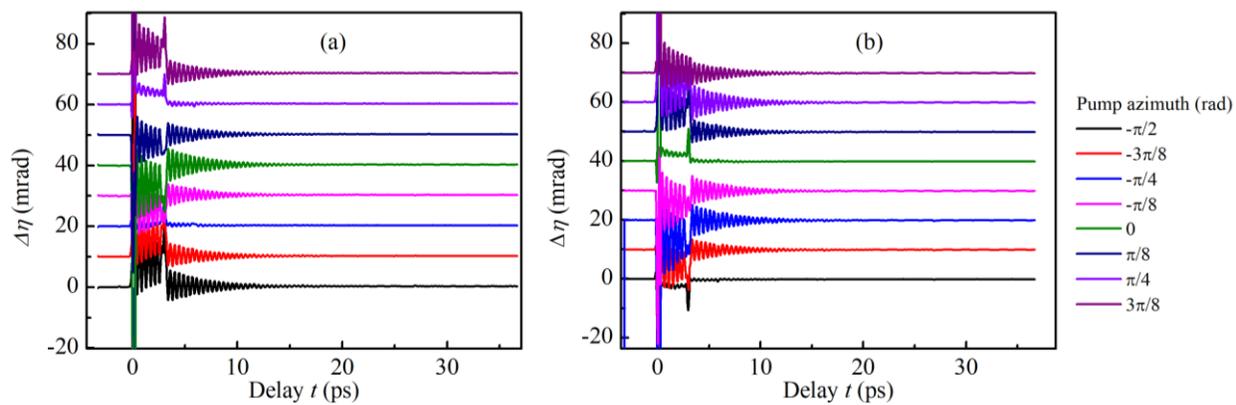

**Fig. S1.** Time-resolved ellipticity changes in transmitted probe polarization of BiFeO$_3$ corresponding to 2.4-THz E phonon mode at several pump azimuth angles in the (a) X- and (b) Y-probe configurations ($T = 300$ K).

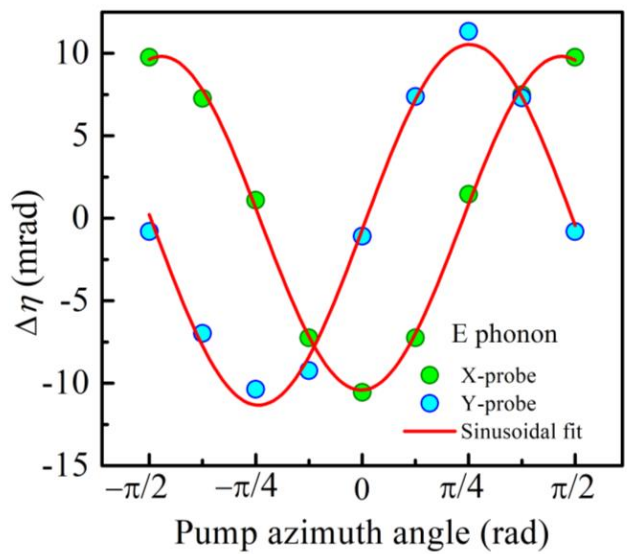

**Fig. S2.** Signed amplitudes of the 2.4-THz mode measured in the X-probe (green circles) and Y-probe (blue circles) configurations as functions of the pump azimuth angle ($T = 300$ K). The solid red curve represents the sinusoidal fit of the experimental data.

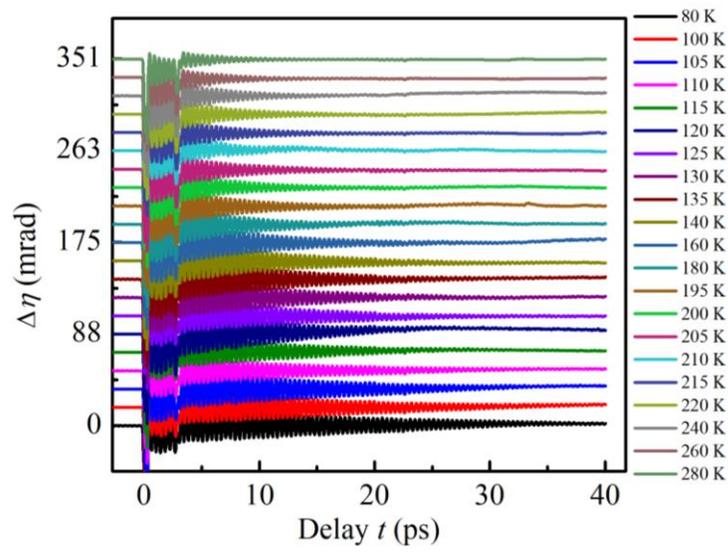

**Fig. S3.** Temporal evolution of the ellipticity changes in transmitted probe polarization of BiFeO$_3$ in the temperature range of 80–280 K up to $t = 40$ ps at $\theta = -\pi/4$ in the Y-probe configuration.

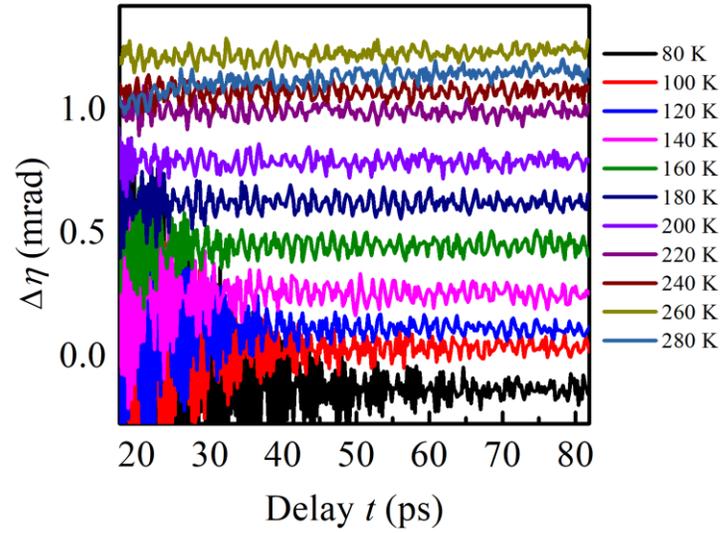

**Fig. S4.** Temporal evolution of the ellipticity changes in transmitted probe polarisation of BiFeO$_3$ in the temperature range of 80–280 K up to $t = 80$ ps at $\theta = -\pi/4$ in the Y-probe configuration. For the initial delay, the few-terahertz oscillation dominates, while at larger delays, the sub-terahertz oscillation dominates.

# Calculation related to the magneto–optical and inverse magneto–optical effects in BiFeO$_3$

**Supplementary Table S2. Tensor components for the 3m symmetry [47].**

| | | | | | | | | | |
|---|---|---|---|---|---|---|---|---|---|
| $k^M$ | $k_1$ | 0 | 0 | | | | | | |
| | 0 | $k_1$ | 0 | | | | | | |
| | 0 | 0 | $k_2$ | | | | | | |
| $k^L$ | 0 | $k_3$ | 0 | | | | | | |
| | $-k_3$ | 0 | 0 | | | | | | |
| | 0 | 0 | 0 | | | | | | |
| $g^{LL}$ | $g_1$ | $g_2$ | $g_3$ | $g_4$ | 0 | 0 | | | |
| | $g_2$ | $g_1$ | $g_3$ | $-g_4$ | 0 | 0 | | | |
| | $g_5$ | $g_5$ | $g_6$ | 0 | 0 | 0 | | | |
| | $g_7$ | $-g_7$ | 0 | $g_8$ | 0 | 0 | | | |
| | 0 | 0 | 0 | 0 | $g_8$ | $g_7$ | | | |
| | 0 | 0 | 0 | 0 | $g_4$ | $g_9$ | | | |
| $p^{ML}$ | 0 | 0 | 0 | 0 | $p_1$ | $p_2$ | 0 | $p_3$ | $p_4$ |
| | 0 | 0 | 0 | 0 | $-p_1$ | $-p_4$ | 0 | $-p_3$ | $-p_2$ |
| | 0 | 0 | 0 | 0 | 0 | $p_5$ | 0 | 0 | $-p_5$ |
| | 0 | 0 | 0 | 0 | $p_6$ | $p_7$ | 0 | $p_8$ | $p_7$ |
| | $-p_7$ | $p_7$ | 0 | $-p_6$ | 0 | 0 | $-p_8$ | 0 | 0 |
| | $-p_9$ | $p_9$ | 0 | $-p_1$ | 0 | 0 | $-p_3$ | 0 | 0 |

The rows and column of a $3 \times 3$ matrix are represented by $\{ij\} = \{yz, zx, xy\}$ and $\{k\} = \{x, y, z\}$. For a $6 \times 6$ matrix, $\{ij\} = \{kl\} = \{xx, yy, zz, yz, zx, xy\}$, while for a $6 \times 9$ matrix, $\{ij\} =$

$\{xx, yy, zz, yz, zx, xy\}$ and $\{kl\} = \{xx, yy, zz, yz, xz, xy, zy, zx, yx\}$. We also define $g_9 = \frac{g_1 - g_2}{2}$ and $p_9 = \frac{p_2 + p_4}{2}$.

We consider $\mathbf{L} = (0, 0, L_z)$, and thus the components of $\epsilon_{ij}(\mathbf{M}, \mathbf{L})$ can be expressed as in Supplementary Table S3.

**Supplementary Table S3**. Dielectric tensor components of BiFeO$_3$ for $\mathbf{L} = (0, 0, L_z)$.

| Tensor element | $\Psi$ mode ($m_x, l_y \neq 0$) | $\Phi$ mode ($m_y, l_x \neq 0$) |
|---|---|---|
| $\epsilon_{xx}$ | $g_4 L_z l_y + p_1 L_z m_x$ | 0 |
| $\epsilon_{yy}$ | $-g_4 L_z l_y - p_1 L_z m_x$ | 0 |
| $\epsilon_{xy}$ | 0 | $g_4 L_z l_x - p_1 L_z m_y$ |
| $\epsilon_{yx}$ | 0 | $g_4 L_z l_x - p_1 L_z m_y$ |

Consequently, we can express the components of $\mathbf{H}^{\text{eff}}$ and $\mathbf{h}^{\text{eff}}$ as

$$H_x^{\text{eff}} = \frac{1}{16\pi} I_0 \delta(t) p_1 L_z \cos(2\theta), \tag{S4}$$

$$H_y^{\text{eff}} = -\frac{1}{16\pi} I_0 \delta(t) p_1 L_z \cos\varphi \sin(2\theta), \tag{S5}$$

$$H_z^{\text{eff}} = 0, \tag{S6}$$

$$h_x^{\text{eff}} = \frac{1}{16\pi} I_0 \delta(t) g_4 L_z \cos\varphi \sin(2\theta), \tag{S7}$$

$$h_y^{\text{eff}} = \frac{1}{16\pi} I_0 \delta(t) g_4 L_z \cos(2\theta), \tag{S8}$$

$$h_z^{\text{eff}} = 0. \tag{S9}$$

For a linearly polarized light ($\varphi = 0$), by using the equations of motion, we express $\mathbf{m}$ and $\mathbf{l}$ as

$$\Delta \mathbf{m}^{\text{lin}} = \frac{1}{32\pi} \gamma I_0 g_4 L_z^2 [\cos(2\theta)\,\hat{x} - \sin(2\theta)\,\hat{y}], \tag{S10}$$

$$\Delta \mathbf{l}^{\text{lin}} = -\frac{1}{32\pi} \gamma I_0 p_1 L_z^2 [\sin(2\theta)\,\hat{x} + \cos(2\theta)\,\hat{y}]. \tag{S11}$$

## Influence of multi-domains on the magneto–optical effects in BiFeO₃

To consider the contradiction observed in the selection rule for the magnon, we assume that for any domain ($n = 0$, 1, and 2) the cycloidal plane is tilted by an angle of $C_n = \frac{2\pi}{3}n$, from the $x$ axis in the $xy$ plane; the new transformed coordinates are denoted as $x'$ and $y'$, as shown in Supplementary Fig. S6.

## Probe light polarization caused by $L_{x'}$

For $L_{x'}$, in $x'$, $y'$, and $z'$ coordinates, under the condition $A_\Psi$ and $A_\Phi \gg 1$, Eqs. (19) and (20) in the main text can be expressed as

$$l_{y'}^{\text{lin}}(t) = \frac{1}{32\pi}\gamma I_0 A_\Psi g_9 L_x^2 \sin(2\theta') \sin(\omega_\Psi t), \tag{S12}$$

$$l_{z'}^{\text{lin}}(t) = \frac{1}{32\pi}\gamma I_0 A_\Phi g_4 L_x^2 \sin(2\theta') \sin(\omega_\Phi t), \tag{S13}$$

where $\theta' = \theta - C_n$, which is the pump azimuth angle in the new coordinate system. Subsequently, in $x$, $y$, and $z$ coordinates, **l** can be expressed as

$$l_x^{\text{lin}}(t) = l_{y'}^{\text{lin}}(t) * (-\sin C_n),\ l_y^{\text{lin}}(t) = l_{y'}^{\text{lin}}(t) * \cos C_n,\ \text{and}\ l_z^{\text{lin}}(t) = l_{z'}^{\text{lin}}(t). \tag{S14}$$

Similarly,

$$L_x = L_{x'} \cos C_n\ \text{and}\ L_y = L_{x'} \sin C_n. \tag{S15}$$

Further, we calculate the components of $\epsilon_{ij}(\mathbf{M}, \mathbf{L})$ by using the second-order magnetization in BiFeO₃ with $\mathbf{L} = (L_x, L_y, 0)$, as shown in Supplementary ~~Note 2~~ Table S4. By using Supplementary Table S4 and substituting **l** and **L** (Eqs. (S14) and (S15)), the polarization changes can be expressed as

$$\Delta\eta_X \propto -g_9 L_y l_y + g_9 L_x l_x + g_4 L_y l_z \propto -A_\Psi(g_9)^2 L_{x'}^3 \sin(2(\theta - C_n)) \sin(2C_n)\sin(\omega_\Psi t) +$$
$$A_\Phi(g_4)^2 L_{x'}^3 \sin(2(\theta - C_n)) \sin(C_n) \sin(\omega_\Phi t), \tag{S16}$$

$$\Delta\eta_Y \propto g_9 L_x l_y + g_9 L_y l_x + g_4 L_x l_z \propto A_\Psi(g_9)^2 L_{x'}^3 \sin(2(\theta - C_n))\cos(2C_n)\sin(\omega_\Psi t) +$$

$$A_\Phi(g_4)^2 L_{x'}^3 \sin(2(\theta - C_n))\cos(C_n)\sin(\omega_\Phi t). \tag{S17}$$

Equations (S16) and (S17) show the change in polarization when the cycloidal plane is tilted at an arbitrary angle from the $x$ axis.

**Probe light polarization caused by $L_z$**

For $L_z$, in $x'$, $y'$, and $z'$ coordinates, under the condition $A_\Psi$ and $A_\Phi \gg 1$, according to Eqs. (25) and (26) in the main text,

$$l_{x'}^{\text{lin}}(t) = \frac{1}{32\pi}\gamma I_0 A_\Phi g_4 L_z^2 \sin(2\theta')\sin(\omega_\Phi t), \tag{S18}$$

$$l_{y'}^{\text{lin}}(t) = \frac{1}{32\pi}\gamma I_0 A_\Psi g_4 L_z^2 \cos(2\theta')\sin(\omega_\Psi t). \tag{S19}$$

Subsequently, in $x$, $y$, and $z$ coordinates, $\mathbf{l}$ can be expressed as

$$l_x^{\text{lin}}(t) = l_{x'}^{\text{lin}}(t) * (\cos C_n) + l_{y'}^{\text{lin}}(t) * (-\sin C_n), \tag{S20}$$

$$l_y^{\text{lin}}(t) = l_{x'}^{\text{lin}}(t) * (\sin C_n) + l_{y'}^{\text{lin}}(t) * (\cos C_n). \tag{S21}$$

Likewise, the changes in probe polarization can be expressed as

$$\Delta\eta_X \propto g_4 L_z l_y \propto A_\Phi(g_4)^2 L_z^3 \sin(2(\theta - C_n))\sin(C_n)\sin(\omega_\Phi t) +$$

$$A_\Psi(g_4)^2 L_z^3 \cos(2(\theta - C_n))\cos(C_n)\sin(\omega_\Psi t), \tag{S22}$$

$$\Delta\eta_Y \propto g_4 L_z l_x \propto A_\Phi(g_4)^2 L_z^3 \sin(2(\theta - C_n))\cos(C_n)\sin(\omega_\Phi t) -$$

$$A_\Psi(g_4)^2 L_z^3 \cos(2(\theta - C_n))\sin(C_n)\sin(\omega_\Psi t). \tag{S23}$$

**Supplementary Table S4**. Components of $\epsilon_{ij}(\mathbf{M}, \mathbf{L})$ in BiFeO$_3$ with $\mathbf{L} = (L_x, L_y, 0)$.

| Tensor element | $m_x, l_x, m_y, l_y, m_z, l_z \neq 0$ |
|---|---|
| $\epsilon_{xx}$ | $g_2 L_y l_y + p_2 L_y m_x + p_3 L_x m_z + g_1 L_x l_x + g_4 L_y l_z + p_4 L_x m_y$ |
| $\epsilon_{yy}$ | $g_1 L_y l_y - p_4 L_y m_x - p_3 L_x m_z + g_2 L_x l_x - g_4 L_y l_z - p_2 L_x m_y$ |
| $\epsilon_{xy}$ | $g_9 L_x l_y - p_9 L_x m_x - p_3 L_y m_z + i k_2 m_z + g_9 L_y l_x + g_4 L_x l_z + p_9 L_y m_y$ |
| $\epsilon_{yx}$ | $g_9 L_x l_y - p_9 L_x m_x - p_3 L_y m_z - i k_2 m_z + g_9 L_y l_x + g_4 L_x l_z + p_9 L_y m_y$ |

**Required summations to consider the contribution of the three domains in BiFeO$_3$**

$$\sum_{1}^{3} \sin(2(\theta - C_n)) \sin(C_n) = \frac{3}{2}\cos(2\theta), \tag{S24}$$

$$\sum_{1}^{3} \sin(2(\theta - C_n)) \sin(2C_n) = -\frac{3}{2}\cos(2\theta), \tag{S25}$$

$$\sum_{1}^{3} \sin(2(\theta - C_n)) \cos(C_n) = \frac{3}{2}\sin(2\theta), \tag{S26}$$

$$\sum_{1}^{3} \sin(2(\theta - C_n)) \cos(2C_n) = -\frac{3}{2}\sin(2\theta), \tag{S27}$$

$$\sum_{1}^{3} \cos(2(\theta - C_n)) \sin(C_n) = -\frac{3}{2}\sin(2\theta), \tag{S28}$$

$$\sum_{1}^{3} \cos(2(\theta - C_n)) \cos(C_n) = \frac{3}{2}\cos(2\theta). \tag{S29}$$

Thus, the change in probe light polarization by $L_x$ under the condition $A_\Psi$ and $A_\Phi \gg 1$ can be expressed as

$$\sum_{n=1}^{3} \Delta\eta_X \propto \frac{3}{2} A_\Psi (g_9)^2 L_x^3 \cos(2\theta) \sin(\omega_\Psi t) + \frac{3}{2} A_\Phi (g_4)^2 L_x^3 \cos(2\theta) \sin(\omega_\Phi t), \tag{S30}$$

$$\sum_{n=1}^{3} \Delta\eta_Y \propto -\frac{3}{2} A_\Psi (g_9)^2 L_x^3 \sin(2\theta) \sin(\omega_\Psi t) + \frac{3}{2} A_\Phi (g_4)^2 L_x^3 \sin(2\theta) \sin(\omega_\Phi t). \tag{S31}$$

Similarly, for $L_z$, under the condition $A_\Psi$ and $A_\Phi \gg 1$,

$$\sum_{n=1}^{3} \Delta\eta_X \propto \frac{3}{2} A_\Psi (g_4)^2 L_z^3 \cos(2\theta) \sin(\omega_\Psi t) + \frac{3}{2} A_\Phi (g_4)^2 L_z^3 \cos(2\theta) \sin(\omega_\Phi t), \tag{S32}$$

$$\sum_{n=1}^{3} \Delta\eta_Y \propto \frac{3}{2} A_\Psi (g_4)^2 L_z^3 \sin(2\theta) \sin(\omega_\Psi t) + \frac{3}{2} A_\Phi (g_4)^2 L_z^3 \sin(2\theta) \sin(\omega_\Phi t). \tag{S33}$$

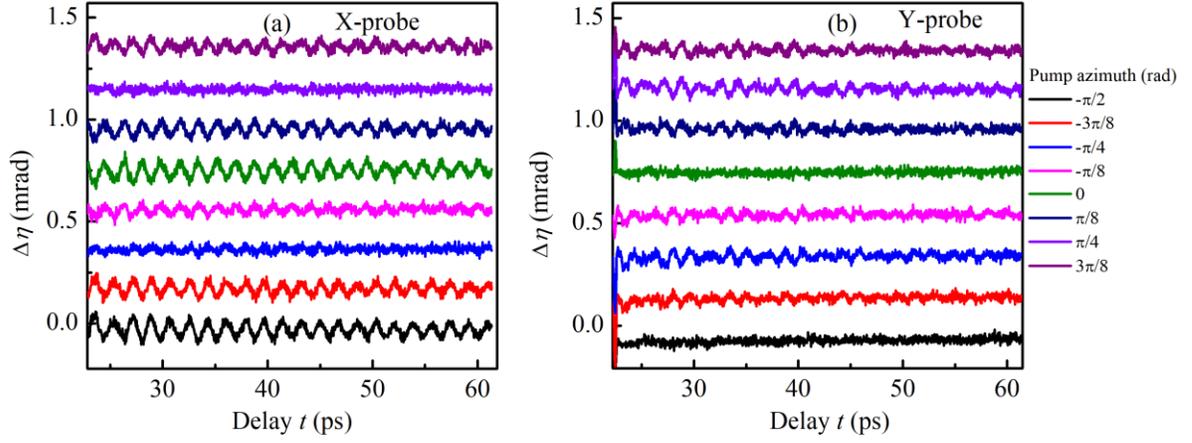

**Fig. S5.** Temporal evolutions of ellipticity changes corresponding to 560-GHz $\Psi_1^{(1)}$ and 740-GHz $\Phi_2^{(1,2)}$ magnon modes at different pump azimuth angles at larger probe delays in the (a) X- and (b) Y-probe configurations ($T = 300$ K).

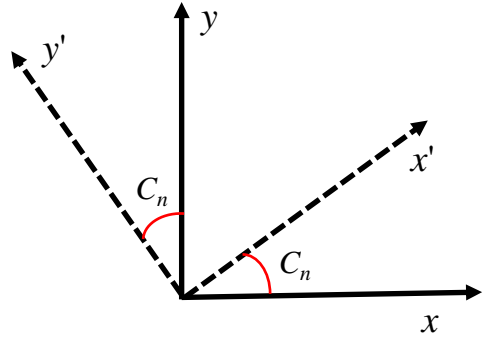

**Fig. S6.** Schematic of the coordinate transformation for any domain $n$, whose cycloidal plane is tilted by an angle of $C_n$ from the $x$ axis in the $xy$ plane. The new transformed coordinates are $x'$ and $y'$.

## Supplementary References